\documentclass[sigconf, nonacm]{acmart}

\usepackage{amsmath}
\usepackage{amsfonts}
\usepackage{graphicx}
\usepackage{verbatim}
\usepackage{siunitx}
\usepackage{subcaption}
\usepackage{csquotes}
\usepackage{mathtools}

\newcommand{\datasetA}{Legal Hate Speech Dataset}


\title{Beyond Imperfect Alternatives with Rulemapping:\texorpdfstring{\\}{ }A Neuro-Symbolic Case Study on Online Hate Speech}

\author{Oskar von Cossel}
\email{cossel@mpipriv.de}
\orcid{0009-0009-2678-4302}

\affiliation{%
  \institution{Max Planck Institute for Comparative and International Private Law}
  \city{Hamburg}
  \country{Germany}
}

\affiliation{%
  \institution{Rulemapping Group}
  \city{Berlin}
  \country{Germany}
}

\begin{document}

\begin{abstract}
Automating legal reasoning forces a choice between imperfect alternatives: symbolic systems offer transparency but struggle with ambiguity, whereas neural systems handle natural language flexibly but lack verifiability. This paper investigates whether a hybrid, neuro-symbolic approach can reconcile this trade-off. We evaluate this architecture in the domain of online content moderation, which serves as a proxy for high-volume legal decision-making such  as mass administrative proceedings. In these settings, operators must assess thousands of cases daily under strict legal standards. Specifically, we examine whether constraining large language models (LLMs) within deterministic symbolic scaffolds improves statute-grounded illegality assessment while preventing \enquote{scope drift} (where LLMs conflate moral offensiveness with legal illegality). We evaluate the neuro-symbolic variant of Rulemapping---a visual logic-tree method that operationalises the classic legal syllogism---on online hate-speech classification under \S~130(1) of the German Criminal Code. Across diverse LLMs, Rulemapping maintains high recall (0.82--0.89) while achieving precision of 0.80--0.86, compared to 0.34--0.49 for unconstrained prompting. Expert-authored symbolic scaffolds thus enable robust legal automation aligned with regulatory requirements for auditability and verifiable decision-making.
\end{abstract}


\maketitle

\section{Introduction}
\label{sec:introduction}

Symbolic and neural approaches to legal reasoning embody opposing
trade-offs. Symbolic systems are transparent, auditable, and offer logical precision but struggle when confronted with unstructured text or statutory ambiguity. Neural systems handle natural language flexibly but remain opaque and, in legal domains, hallucinate at alarming rates~\cite{dahlLargeLegalFictions2024,mageshHallucinationFreeAssessingReliability2025}.

Early symbolic systems showed that domain logic can, to an extent, be encoded as executable computer programs: MYCIN~\cite{shortliffeMycinKnowledgeBasedComputer1977} demonstrated this in medicine in 1977, and, in law, the Imperial College Group (ICG) subsequently realised a logic-program representation of the British Nationality Act in Prolog~\cite{sergotBritishNationalityAct1986}---an achievement that also provoked intensive critique~\cite{molesLogicProgrammingAssessment1991}. A central observation was that symbolic reasoning cannot adequately accommodate the open texture---vagueness and context-dependency---inherent in legal language~\cite{bench-caponRulebasedRepresentationOpen1988}. A further point of critique was that the ICG did not consult domain experts sufficiently, resulting in low information density~\cite{molesLogicProgrammingAssessment1991}. This illustrates a second key problem of symbolic approaches: experts must provide knowledge top-down, a process that becomes especially demanding when they must understand the syntax and mathematical logic underlying logic programming languages, a challenge commonly referred to as the knowledge-acquisition bottleneck.

Over the past two decades, advances in neural architectures---word-to-vector embeddings~\cite{mikolovDistributedRepresentationsWords2013}, transformers~\cite{vaswaniAttentionAllYou2017}, and culminating in ever-improving LLMs---have largely bypassed these constraints. In computational law, neural approaches are rapidly growing or even dominating, notably for summarisation tasks, argument mining, and classification tasks such as legal judgment prediction and question answering (see~\cite{medvedevaLegalJudgmentPrediction2023,ariaiNaturalLanguageProcessing2025} for overviews). This shift represents a substantial move away from systems that rely primarily on top-down expert knowledge, toward a bottom-up data-centric regime where ontologies, code libraries, and datasets (such as~\cite{bommaritoKL3MDataProject2025,fanLEXamBenchmarkingLegal2025}) constitute the primary sources of knowledge. However, this flexibility comes at a steep cost. Empirical studies document that LLMs hallucinate citations and precedents in U.S. case law in 58--88\% of queries~\cite{dahlLargeLegalFictions2024}, and even specialised AI-powered legal research tools likewise exhibit notable hallucination rates of up to 33\%~\cite{mageshHallucinationFreeAssessingReliability2025}. Moreover, their opacity often violates the transparency and explainability requirements set out in the EU AI Act and GDPR for high-stakes decision-making~\cite{paniguttiRoleExplainableAI2023}. The field thus faces a fundamental tension, as neither approach, in isolation, overcomes the trade-off between the semantic flexibility of neural models and the transparency and determinism of symbolic systems. 

To move beyond these imperfect alternatives, research increasingly turns to hybrid neuro-symbolic architectures (e.g.~\cite{mumfordReasoningLegalCases2022,zhangSyLeRFrameworkExplicit2025,kantEquitableAccessJustice2024}). By constraining neural components within symbolic scaffolding, these systems seek to restore determinism and auditability while preserving the linguistic capabilities of modern NLP.

We investigate a hybrid approach grounded in this direction: the neuro-symbolic variant of Rulemapping. Rulemapping itself was originally developed as a purely symbolic framework~\cite[p. 348]{breidenbachRechtshandbuchLegalTech2021} with a focus on practicability and accessibility, and has seen sustained commercial deployment for more than two decades across German courts, law firms, and administrative agencies. It addresses the knowledge-acquisition bottleneck through visual, expert-driven design: legal professionals encode statutory logic directly using tree structures, without requiring mastery of logic programming syntax. To mitigate the hallucination and opacity risks inherent in purely neural systems, the neuro-symbolic variant confines LLM assessments to specific open-textured elements, retaining a deterministic symbolic scaffold for the overall reasoning process. By strictly constraining LLMs to these specific sub-tasks, this variant aims to restore the auditability and determinism of symbolic systems while leveraging the semantic understanding of current models.

We evaluate this neuro-symbolic variant on the task of online hate-speech classification under \S~130(1) of the German Criminal Code (Strafgesetzbuch; StGB), which governs incitement to hatred. This domain presents a direct conflict between the industrial scale of content moderation and the doctrinal nuance of criminal law. The EU Digital Services Act requires platforms and their Article~21 dispute-settlement bodies to conduct complex, context-sensitive legal assessments across massive volumes of user content under strict timelines~\cite{ederMakingSystemicRisk2024}. Such conditions demand automation that can reconcile procedural speed with the depth of reasoning expected of judicial decision-making.

Our work makes two key contributions. First, we empirically validate our core hypothesis: that constraining neural models via deterministic tree composition improves reliability in legal judgment classification. Using Rulemapping as an exemplar of this architecture, we validate this claim on \S~130(1) StGB, a norm of high interpretive complexity. The results show that neuro-symbolic decomposition operationalises the syllogistic reasoning of civil-law jurisdictions, suggesting applicability to rule-intensive domains such as administrative law. Through systematic benchmarking against LLM-only baselines on a legally annotated dataset, we show that this approach yields superior precision-recall trade-offs (e.g., Rulemapping precision of $0.86$ vs. baseline $0.49$ for Mistral, see Table~\ref{tab:bench-zufall-lay}) while preserving recall and reducing the hallucination and over-flagging risks inherent in free-form generation.

Second, we show how this method supports real-world application by legal experts, mitigating the knowledge-acquisition bottleneck through visualisation and logical simplicity. We demonstrate how statutory elements are encoded in visual, Boolean logic trees, how contextual materials are selected and scoped to specific judgments, and where neural assistance is appropriate. By keeping humans in the loop via transparent, inspectable representations, Rulemapping also aligns with AI Act and GDPR transparency demands.

The remainder of this paper is structured as follows. Section~\ref{sec:related-work} situates our approach within the broader landscape of neuro-symbolic legal AI and visualisation techniques. Section~\ref{sec:rulemapping} explains Rulemapping and its neuro-symbolic variant. We present the methodology for our case-study in Section~\ref{sec:case-study} and report empirical results against lay and expert reference annotations in Section~\ref{sec:results}. Section~\ref{sec:discussion} discusses these findings, followed by an analysis of limitations and legal considerations in Section~\ref{sec:limitations}. We offer concluding remarks in Section~\ref{sec:conclusion} and propose directions for future research in Section~\ref{sec:future-work}.

\section{Related Work}
\label{sec:related-work}

The problems mentioned in Section~\ref{sec:introduction} have been tackled from different directions. We review visualisation techniques and neuro-symbolic architectures---both directly relevant to Rulemapping's design---before detailing the method itself.

\subsection{Visualisation Techniques for Accessible Knowledge Formalisation}

Visualisation addresses the knowledge-acquisition bottleneck by making symbolic formalisms accessible to non-technical legal professionals~\cite{streebTaskBasedVisualInteractive2022,nguyenLawBinaryTree2022}. Early research focused on argumentation diagramming; tools such as Araucaria~\cite{reedAraucariaSoftwareArgument2004} and DefLog~\cite{verheijDefLogLogicalInterpretation2003} use structured logic to represent arguments and defeat conditions. Beyond these early approaches, contemporary systems pursue alternative strategies. For example,~\cite{nguyenLawBinaryTree2022} employs a binary-tree structure to represent legal norms as conditional decision hierarchies, whereas~\cite{mclachlanLawmapsEnablingLegal2022} uses flowcharts to visualise the procedural flow and decision logic embedded in legislation and legal practice. A useful overview of specific visualisation techniques in the legal literature is provided in~\cite{mclachlanVisualisationLawLegal2021}, within which Rulemapping is best categorised as a concept-map approach. Recent work in Visual Analytics likewise employs concept-map-style visualisations, among other techniques, to externalise tacit legal knowledge in jurisprudence~\cite{furstChallengesOpportunitiesVisual2025}. Rulemapping, however, distinguishes itself by focusing on the granular logical decomposition of statutory elements. Taken together, these approaches show that domain experts can participate directly in knowledge representation when the underlying formalism is visually grounded rather than expressed in code.

\subsection{The Symbolic[Neuro] Integration Pattern}
\label{sec:related-symbolic-neuro}

Neuro-symbolic architectures, as described in Section~\ref{sec:introduction}, cover a wide range of ways to integrate symbolic reasoning with LLMs. Kautz's taxonomy~\cite{kautzThirdAISummer2022} offers a principled framework for organising these integration patterns. Among the architectural patterns Kautz identifies, we focus on the Symbolic[Neuro] pattern, in which a symbolic system orchestrates problem decomposition and delegates narrowly scoped subroutines to neural modules. AlphaGo~\cite{silverMasteringGameGo2016} exemplifies this architecture: policy and value neural networks guide a Monte Carlo Tree Search algorithm within a symbolic search space.

Similar Symbolic[Neuro] methods have emerged in legal reasoning. The ADF-ML system described in~\cite{mumfordReasoningLegalCases2022} learns factor ascriptions with factor-specific hierarchical BERT models and composes them in an Abstract Dialectical Framework (ADF) to resolve issues and outcomes for Article~6 ECHR cases, operationalising precedent-sensitive argumentation via learned leaf classifiers. A Symbolic[Neuro] Prolog pipeline~\cite{kantRobustLegalReasoning2025} uses LLMs to map text to ground facts that a symbolic engine then executes. SyLeR~\cite{zhangSyLeRFrameworkExplicit2025} follows a similar pattern but uses tree-structured retrieval over statutes and precedents combined with explicitly structured syllogistic reasoning. Their shared design goal is to constrain neural components within symbolic scaffolds that maintain transparency and improve complex reasoning.

When comparing these Symbolic[Neuro] approaches through a three-step workflow---(1) logical scaffold specification, (2) processing and evaluation of natural-language inputs via LLMs, and (3) symbolic computation---the first step reveals a central design trade-off. In some systems, the scaffold itself is automated or partially learned; for instance, SyLeR~\cite{zhangSyLeRFrameworkExplicit2025} derives its reasoning structure dynamically from retrieval and learned classifiers. By contrast, others rely on fixed, expert-authored scaffolds. ADF-ML~\cite{mumfordReasoningLegalCases2022} employs an expert-defined ADF to represent domain knowledge, while~\cite{kantRobustLegalReasoning2025} encodes expert knowledge as logic programs. However, both approaches require translating domain knowledge into technical formalisms, which effectively reintroduces the knowledge-acquisition bottleneck for experts lacking technical expertise.

Rulemapping instead deliberately assigns scaffold construction to domain experts, who design the logical structure from statutes and legal literature using visual Rulemaps, rather than code. This choice requires expert effort upfront but reduces dependence on task-specific training data and maintains high transparency, as labels and structure remain easily inspectable and modifiable even for practitioners without training in logic programming. By combining accessible visual representation with constrained neuro-symbolic integration, Rulemapping synthesises the two research trajectories reviewed above.

\section{Rulemapping}
\label{sec:rulemapping}

We now outline Rulemapping and illustrate the approach by reference to \S~130(1)~no.~1 StGB with a running example (Section \ref{sec:method}). We then specify the neuro-symbolic variant (Section \ref{sec:neuro-symbolic}) in which a symbolic controller preserves the Rulemap's logical scaffold while delegating vague and open-textured leaf assessments to LLMs, with the symbolic controller defining both their scope and permissible outputs.

\subsection{Symbolic Method}
\label{sec:method}

Rulemapping formalises legal reasoning through structured logical decomposition by breaking down legal norms into discrete constituent elements and representing the logical relationships between these elements using standard propositional logic operators in a distinct visualisation framework~\cite[p. 348]{breidenbachRechtshandbuchLegalTech2021}. It is best characterised as a primarily rule-based approach to the judgment phase of legal reasoning grounded in syllogistic decomposition, rather than a case-based paradigm. The methodology builds upon the judicial syllogism---the deductive reasoning pattern that applies legal rules to factual situations. This approach finds expression in various national legal traditions: Germany's \textit{Subsumtionstechnik} represents one systematic implementation of syllogistic reasoning in legal analysis~\cite[p. 92]{larenzMethodenlehreRechtswissenschaft1995}, while common-law jurisdictions employ similar deductive structures through frameworks like the IRAC method (Issue-Rule-Application-Conclusion). The resulting formal structure permits both manual analysis by legal practitioners and computational processing through automated systems.

Yet law remains fundamentally a text-based discipline, where juristic knowledge is embedded in statutes, court decisions, legal publications, and commentaries. A key concept of Rulemapping is translating these text-bound rules into visualised expert trees with fixed logical structure, promoting transparency and practitioner accessibility for complex regulatory logic. This contrasts with sentence-based approaches such as PROLEG~\cite{satohPROLEGPracticalLegal2023}, which requires legal knowledge to be authored in a programming formalism, reintroducing the knowledge-acquisition bottleneck that visualisation aimed to mitigate.

Rulemapping implements this translation via rooted trees of logical nodes in parent--child relationships, with each parent computing a logical operator over its children: AND ($\land$), OR ($\lor$), XOR ($\veebar$), each optionally negated ($\lnot$)~\cite[p. 348]{breidenbachRechtshandbuchLegalTech2021}. Unlike classic expert trees or flowchart logics such as~\cite{mclachlanLawmapsEnablingLegal2022}, evaluation proceeds bottom-up from the leaves to the root. Building a Rulemap generally begins by phrasing a top-level question (root node) and refining it via its branches into constituent sub-questions at lower levels (leaf nodes). Boolean child truth values then propagate upward according to the specified connectives to determine each parent's truth value and ultimately the truth value of the root node.

We illustrate Rulemap development using \S~130(1) no.~1 StGB (Incitement to hatred), which states:

\begin{quote}
Whoever, in a \emph{manner suited to causing a disturbance of the public peace},
\sloppy
\begin{enumerate}
  \item \emph{incites hatred} against a \emph{national, racial, religious group} or a \emph{group defined by their ethnic origin}, against \emph{sections of the population} or \emph{individuals on account of their belonging to one of the aforementioned groups or sections of the population}, or \emph{calls for violent or arbitrary measures against them}
  \item\relax [...]
\end{enumerate}

incurs a penalty of imprisonment for a term of between three months and five years.\footnote{Italics were added for emphasis. The quotation is from the unofficial translation of \S~130 StGB: \url{https://www.gesetze-im-internet.de/englisch_stgb/englisch_stgb.html\#p1368}; official German version: \url{https://www.gesetze-im-internet.de/stgb/__130.html}}
\end{quote}

Beyond \S~130(1) StGB, the statute contains seven further subsections that establish five additional distinct case variants~\cite{anstotz130StGBVolksverhetzung2025}. For purposes of illustration, we restrict our analysis to the variant cited in subsection~(1) no.~1, which---in contrast to many other provisions---explicitly enumerates all elements of the offence within its statutory text.

\begin{figure}[!ht]
  \centering
  \includegraphics[width=\columnwidth]{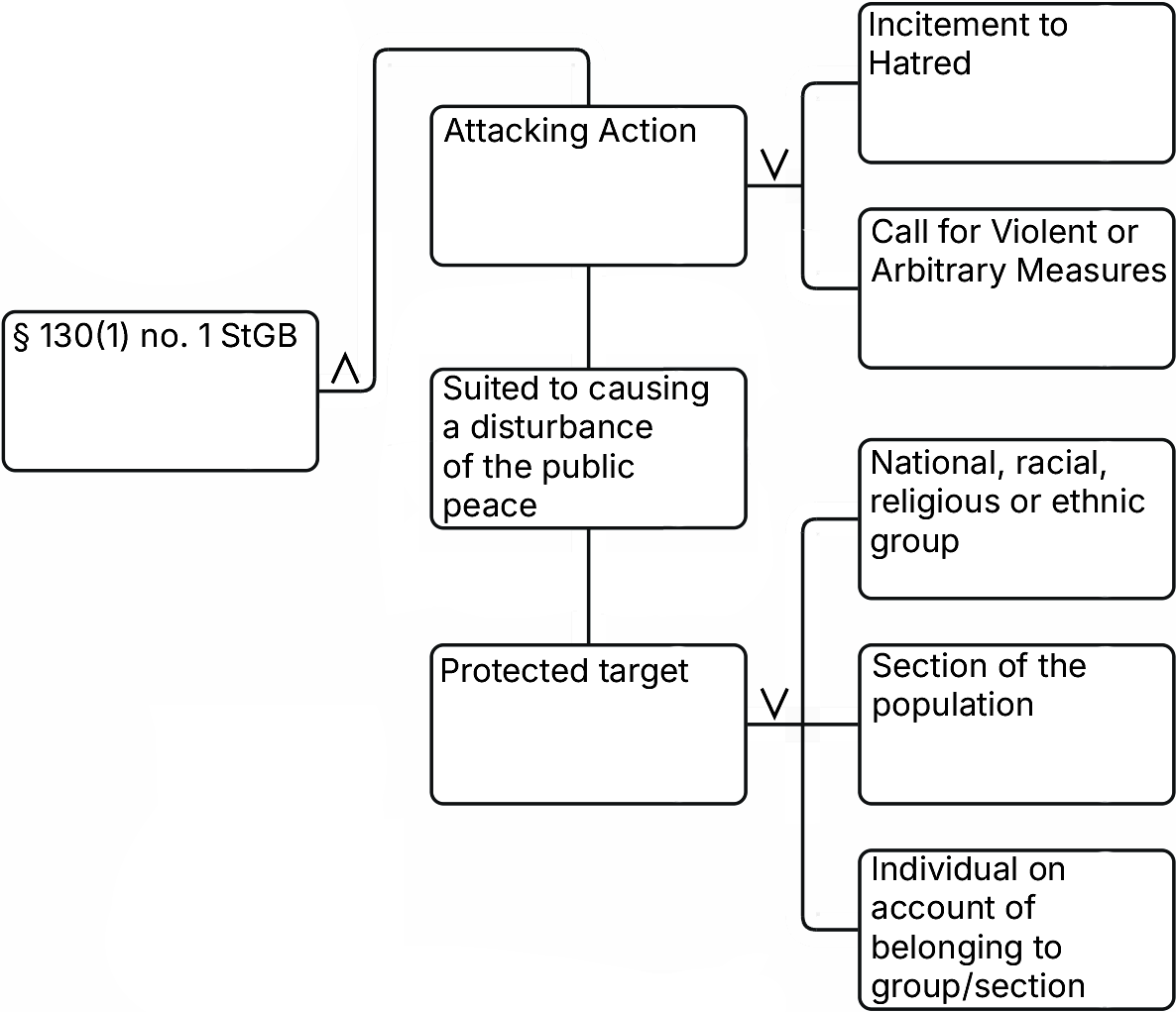}
    \Description[Rulemap of \S~130(1)~no.~1~StGB.]{Description in subsequent text.}
  \caption[Rulemap of \S~130(1) no. 1 StGB]{
    Rulemap of \S~130(1)~no.~1~StGB (Incitement to hatred); left-to-right flow from root to leaf nodes, discussed in Section~\ref{sec:method}. Rulemap built using publicly available logic-tree visualisation software\footnotemark\ and slightly edited for readability.
  }\label{fig:rulemap130}
\end{figure}

\footnotetext{The Rulemap was constructed using the Rulemap builder at \url{https://builder.rulemapping.org/}. We made slight changes for readability.}

When represented as a Rulemap (see Figure \ref{fig:rulemap130}), establishing \S~130(1) no.~1 StGB requires satisfying the \textit{root node} through three nodes connected by \( \land \): an \textit{attacking action} (parent node), \textit{suited to causing a disturbance of the public peace} (leaf node), and a \textit{protected target} (parent node). The \textit{attacking action} node is satisfied via \( \lor \) when the expression constitutes \textit{incitement to hatred} (leaf node) or a \textit{call for violent or arbitrary measures} (leaf node); these alternatives are not necessarily exclusive, so \( \veebar \) does not apply. The \textit{protected target} node is satisfied via \( \lor \) when the addressee is a \textit{national, racial, religious or ethnic group} (leaf node), a \textit{section of the population} (leaf node), or an \textit{individual on account of belonging to such a group or section} (leaf node). Thus, the \textit{root node} is established when at least one \textit{attacking action} leaf and one \textit{protected target} leaf are present and the suitability leaf is fulfilled.

The following propositional logic transcription formalises the visual structure for clarity:
\[
\begin{aligned}
\textit{\S~130(1) no.~1}\;\coloneqq{}& \textit{Attacking action}\\
&\!\land \textit{Suited to disturb the public peace}\\
&\!\land \textit{Protected target}
\\[4pt]
\textit{Attacking action}\;\coloneqq{}& \textit{Incitement to Hatred}\\
&\!\lor \textit{Call for violent or arbitrary measures}
\\[4pt]
\textit{Protected target}\;\coloneqq{}& \textit{Protected Group}\\
&\!\lor \textit{Section of population}\\
&\!\lor \textit{Individual}
\end{aligned}
\]

To further illustrate Rulemapping, we consider a contested boundary case that exposes the interpretive challenges inherent in statutory elements. This example uses a social media post from the \datasetA~\cite{zufallLegalApproachHate2022} (translated from German by the authors): \enquote{\textit{The degenerate subhuman race of German potatoes deserves nothing less than to be wiped out and replaced.}}

Can this statement satisfy \S~130(1) no.~1 StGB?
The \textit{attacking action} parent node becomes true through its call-for-violent-measures leaf. Contrary to what one might expect, \S~130 protects not only minorities but also---in the German context---majority populations~\cite[p.~381]{mitschVolksverhetzungGegenDeutsche2011}. In this case, \enquote{German potatoes} is a derogatory slang term for ethnic Germans, drawing on both the potato's cultural and culinary association and the established pejorative use of the German word \enquote{Kartoffel} (literally: \enquote{potato}). Thus the \textit{protected target} parent node is satisfied on two grounds: Germans constitute a \enquote{national group} as explicitly enumerated in the statute, and they qualify as a \enquote{section of the population} because they comprise only approximately 90\% of Germany's residents, making them numerically a distinguishable section rather than the entirety~\cite[p.~381]{mitschVolksverhetzungGegenDeutsche2011}. Crucially, however, the \textit{suitability of causing a disturbance of the public peace} leaf fails. Despite being a protected group, Germans form the demographic majority and thus cannot experience the systemic vulnerability and realistic threat of exclusion that the disturbance of the public peace element requires~\cite[no.~3]{sternberg-lieben130StGBVolksverhetzung2025}. Since the root requires all three conjuncts (\( \land \)), a single failing leaf defeats the entire condition: \( \top \land \bot \land \top \Longleftrightarrow \bot \). The statement thus does not constitute punishable incitement under \S~130(1) no.1 StGB, despite satisfying two of three statutory elements.

\subsection{Neuro-Symbolic Variant}
\label{sec:neuro-symbolic}

Rulemapping performs strongly where rules are determinate and quantifiable, yet it needs non-symbolic support to handle open-textured assessments---vague or ambiguous elements requiring contextual evaluation~\cite{hartConceptLaw1961,guittonChallengeOpentextureLaw2025}---and to interpret unstructured inputs that symbolic systems cannot map without predefined schemas.

These limitations can be illustrated by the social-media example discussed in Section \ref{sec:method}. The post is an unstructured text, and a purely symbolic process struggles to parse it reliably. In addition, the statutory elements in the example are open-textured to varying degrees. At the lower end, the protected groups are exhaustively listed; but even here, as already explained above, the boundaries remain fluid~\cite[no.~3]{sternberg-lieben130StGBVolksverhetzung2025}. The element \enquote{Section of the Population} is even less determinate. At the upper end of open-texture lies the \enquote{Attacking Action}, which even triggers constitutional balancing: the protected public interest in public peace and the human dignity of the targeted individuals under Art. 1 of the German Constitution (Grundgesetz; GG) must be weighed against the statement author's freedom of expression under Art. 5(1) sentence 1 GG~\cite[no.~5]{sternberg-lieben130StGBVolksverhetzung2025}, which makes the outcome even more context-dependent.

In response, Rulemapping employs a neuro-symbolic design that retains the Rulemap's logical scaffold and delegates leaves that require interpretation of unstructured or open-textured content to a constrained LLM operating under curated, leaf-specific context. During construction, the expert author determines for each leaf whether it requires LLM interpretation of open-textured content or whether it can be computed symbolically from structured records and deterministic algorithms (e.g., limitation periods from dates). This design minimises LLM usage in order to preserve determinism. Where input data arrives primarily as unstructured text, such as in the running hate-speech example, this may result in all leaves being evaluated via LLMs. Yet also in this configuration, the symbolic layer strictly governs decomposition and composition as explained in Section~\ref{sec:rulemapping}, ensuring the LLM operates only as a bounded subroutine. Each LLM-evaluated leaf returns a Boolean truth value through structured prompt instructions that request an explicit binary response (e.g., \enquote{yes} or \enquote{no}), which the system parses and validates before propagating the result upward through the tree's logical operators.

This approach prioritises contextual materials that decision-makers actually use. In Germany's civil-law system, legislation is the primary source of law, and courts routinely consult article-by-article legal commentaries that situate provisions within the code and synthesise doctrine and case law; although persuasive rather than binding, these works are central to structuring judicial reasoning along the relevant chains of provisions~\cite{djeffalCommentaryCommentariesWissenschaftsrat2014}. To operationalise this principle, domain experts curate and store context for each statutory element at its corresponding leaf node. The creation of this context follows a structured, deductive logic analogous to the definitional analysis in the \textit{Subsumtionstechnik} or the Rule step in IRAC (see Section~\ref{sec:method}), ensuring materials are as brief as possible yet as extensive as necessary. Rather than incorporating large, unstructured text segments, this method favors a concise synthesis of rules drawn from selected landmark decisions, administrative guidance, and illustrative general examples, mirroring the curated logic of a formal legal opinion.

Modular context curation at leaf nodes facilitates responsiveness to doctrinal shifts. When new precedents clarify an open-textured element (e.g., the threshold for \enquote{public peace}), experts update that specific leaf's context material directly. Because the scaffold is deterministic, these local updates propagate predictably, allowing the system's legal knowledge to evolve iteratively alongside case law.

\section{The Hate Speech Case Study}
\label{sec:case-study}

To validate the neuro-symbolic Rulemapping approach empirically, we benchmark it against prompt-based baselines on online hate-speech classification under \S~130(1) StGB, introduced as a running example in Section~\ref{sec:method}. All methods are evaluated on a separate human-annotated dataset that provides the gold-standard reference and is described in Section~\ref{sec:dataset}. The experimental series targets a single binary endpoint---statutory illegality under \S~130(1) StGB---using fixed textual inputs (social-media posts) and human reference annotations.

\subsection{Dataset and Experimental Design}
\label{sec:dataset}

The experiments use the \datasetA~\cite{zufallLegalApproachHate2022}, which comprises 1,000 German-language social media posts annotated for violations of EU Framework Decision 2008/913/JHA as implemented in \S~130(1) StGB in Germany. Each post received binary labels on 15 statutory elements from two paid lay annotators, with roughly 10\% of cases subsequently reviewed by a legal expert. The annotators received extensive context and guidelines~\cite{zufallLegalApproachHate2022}. The dataset creators sourced the posts from court decisions, web searches, and the GermEval-2019 corpus, supplemented by synthetic examples they generated. A binary punishability label, derived from the 15 element-level annotations following the logic in~\cite[p.~55]{zufallLegalApproachHate2022}, serves as the gold standard reference. The data shows strong class imbalance, with only a small minority of posts satisfying \S~130(1) StGB, a distribution typical of hate-speech datasets. For all experiments, where supported, temperature is set to $0$, top-$p$ to $0.01$, and the random seed to $640$. We selected these decoding parameters to minimise stochastic variance and ensure that performance differences result from the prompting strategies rather than random generation artifacts.

\subsubsection{Rulemapping Method}

The experiments employ the neuro-symbolic Rulemapping method introduced in Section~\ref{sec:neuro-symbolic}, using a slightly more fine-grained tree for \S~130(1) StGB than in the example. The leaf nodes delegate bounded judgments to an LLM with element-specific context drawn from legal commentaries and court decisions, and the symbolic controller aggregates the resulting truth values via propositional logic operators to obtain the final label. The Rulemap was authored by a legal researcher with experience in formal rule modelling but no prior specialisation in \S~130 StGB. Consequently, the initial construction phase---comprising statutory analysis, literature review, logical decomposition, and context curation to derive the necessary domain knowledge---required approximately three hours. This scaffold was subsequently refined against a pilot set of approximately 50 hate-speech cases outside the benchmark dataset, a process that required an additional four to five hours of expert effort.

\subsubsection{Long-Context Baseline}

The long-context baseline is a purely LLM-based approach that evaluates zero-shot chain-of-thought prompting~\cite{weiChainofthoughtPromptingElicits2022} with the same stepwise reasoning scaffold and statutory context as Rulemapping. This prompting style has been shown to improve performance on multi-step reasoning benchmarks in the zero-shot regime by instructing models to \enquote{think step by step} along a series of intermediate reasoning steps, yielding substantial accuracy gains on arithmetic, symbolic, and logical tasks~\cite{weiChainofthoughtPromptingElicits2022}. 

The prompt instructs the model to apply the German legal opinion style (\textit{Gutachtenstil}) and evaluate each statutory element sequentially, then produce a binary classification under \S~130(1) StGB. It provides the full statutory text, detailed definitions for each element, and explicitly describes the logical relationships between them (e.g., which elements must be satisfied cumulatively versus alternatively). However, unlike Rulemapping, this logical structure is provided only as natural language instruction. Inference proceeds entirely via free-form generation with no explicit computational constraints---meaning the decomposition of statutory elements serves only as a soft reasoning guide, not as a deterministic scaffold that forces the final label to match the intermediate findings.

\subsubsection{Zero-Context Baseline}

The zero-context baseline is a purely LLM-based approach that evaluates instruction-only classification. The model receives a minimal task description and the raw social-media post, with no auxiliary definitions, examples, rationales, or external context provided. This setup probes out-of-the-box capabilities for statute application without in-context scaffolding, following common zero-shot practice in legal NLP.

The system prompt reads: \texttt{Du musst Kommentare, die in sozialen Medien gepostet wurden auf deren Strafbarkeit gemäß \S 130 Abs. 1 StGB (Volksverhetzung) bewerten. Du hast nur 2 Antwortmöglichkeiten: y für yes->Strafbar; n für no-> NICHT strafbar.}\footnote{English translation: \enquote{You must assess comments posted on social media for their punishability under \S~130(1) StGB (incitement to hatred). You have only 2 answer options: y for yes->punishable; n for no->NOT punishable.}} The user prompt then provides the post text.

\subsubsection{LLM Selection}

LLM selection targets three performance tiers to assess how performance varies across model scales and architectures. The reasoning-optimised tier includes GPT-o1~\cite{openaiOpenAIO1System2024}, which employs extended inference-time computation and internal chain-of-thought before generating outputs, and GPT-5 mini~\cite{singhOpenAIGPT5System2025}, a smaller reasoning-optimised model designed for cost and latency efficiency. The large general-purpose tier comprises GPT-4o~\cite{openaiGPT4oSystemCard2024} and Deepseek V3~\cite{deepseek-aiDeepSeekV3TechnicalReport2025}, both frontier models that represent the upper bound of general legal and reasoning performance available at the time of this study. The mid-scale tier includes OSS 120B~\cite{openaiGptoss120bGptoss20bModel2025} and Mistral Large~\cite{mistralaiteamPixtralLargeMistral2024}, both strong open-weight models that allow testing whether Rulemapping's symbolic decomposition can compensate for reduced capacity relative to frontier systems. 

\subsection{Related Benchmarks and Datasets}

While evaluation frameworks for LLMs continue to expand, datasets designed for legal reasoning remain scarce, particularly for German law. Only approximately 1\% of German court decisions are published~\cite{hamannTransparenzJustizStagnation2021}, constraining the empirical foundation for legal NLP evaluation. A benchmark that spans German private law very broadly in an open question format~\cite{conradsJuristischeProblemlosungMit2025} shows that state-of-the-art LLMs progressively narrow the gap with human performance but remain strongest in assistive roles rather than autonomous judgment.

LEXam~\cite{fanLEXamBenchmarkingLegal2025} provides a further broad reference point through an open-source bilingual dataset of English and German legal question-answer pairs, scored via an LLM-as-a-judge protocol. On the open-questions subset, model performance on the LEXam benchmark for the models used here is as follows: GPT-5 mini (60.32/100), GPT-4o (56.93/100), Deepseek V3 (52.53/100), and GPT-OSS-120B (51.74/100), while Mistral Large and GPT-o1 are not included.\footnote{See \url{https://lexam-benchmark.github.io/}} LEXam also reports systematically lower performance on German items than on English ones, aligning with the general scarcity and language-specific difficulty of German-law evaluation~\cite{fanLEXamBenchmarkingLegal2025}.

In the domain of hate speech, the DeTox Dataset~\cite{demusDeToxComprehensiveDataset2022} offers a large-scale alternative to the one we used, with 10,278 German-language tweets annotated for general toxicity characteristics and specific criminal offenses (including \S~130 StGB). Concerning classification methodology, prior work shows that prompting strategy substantially affects outcomes. In~\cite{guoInvestigationLargeLanguage2023}, different prompting approaches were tested on English hate-speech datasets. The study demonstrated that different prompting approaches yield significant performance variations, with chain-of-thought reasoning showing the most promise. 

Closest to the present setting, a benchmark on the \datasetA~used in this paper evaluates a subset of 150 cases and compares prompting strategies for classification under \S~130(1) StGB~\cite{ludwigConditioningLargeLanguage2025}. It reports that providing more context can underperform simpler prompting variants, indicating that additional contextual scaffolding does not reliably improve performance under this dataset's annotation scheme~\cite{ludwigConditioningLargeLanguage2025}.

\section{Case Study Results}
\label{sec:results}

We evaluate three approaches: Rulemapping as the primary method, and two non-neuro-symbolic LLM-only baselines---Long-Context prompting (zero-shot chain-of-thought with curated materials) and Zero-Context prompting (instruction-only classification)---against (i) the lay-consensus reference, defined as the subset of 868 instances on which both lay annotators agreed (excluding 132 disagreements), and (ii) a stratified expert subset comprising 102 instances annotated by a single domain expert.

Because the positive class is rare, Accuracy can remain high even when a system fails to identify positive instances. We therefore treat Precision and Recall as the primary metrics and report Accuracy only as a descriptive complement.

Precision measures the proportion of predicted positives that are correct (penalising false positives), whereas Recall measures the proportion of true positives recovered (penalising false negatives). Formally, 

\[
\text{Precision}\ (P)=\frac{\mathrm{TP}}{\mathrm{TP}+\mathrm{FP}},\qquad
\text{Recall}\ (R)=\frac{\mathrm{TP}}{\mathrm{TP}+\mathrm{FN}}.
\]

In settings with heavy class imbalance, reporting both \(P\) and \(R\) makes method trade-offs explicit: high \(P\) with low \(R\) indicates conservative behavior (few false positives but many misses), whereas high \(R\) with low \(P\) indicates aggressive flagging (few misses but many false alarms). This choice enables a transparent accounting of false-positive and false-negative behavior across methods rather than collapsing these error types into a single-number summary (e.g., $F_1$).

The accompanying repository\footnote{Supplementary material: \url{https://github.com/ovcsl/rm-hs-bench}} reports, for every configuration, the full confusion matrix ($TP$,$FP$,$TN$,$FN$) together with $F_1$, $F_2$, Balanced Accuracy and Cohen's $\kappa$ enabling precise comparability with related work and full reproducibility of all summary statistics.

\subsection{Results (Lay-Consensus Reference)}
\sloppy
Results for the lay-consensus reference are shown in Table~\ref{tab:bench-zufall-lay}, grouped by method and model. Rulemapping and prompt-only baselines show inverse precision-recall profiles. Prompt-only methods skew heavily toward recall (typically 0.89--1.00) but suffer from low precision, generally staying below 0.50 (e.g., Mistral Large Long-Context reaches 0.49, others sit between 0.28--0.37). Rulemapping's top performers maintain high recall (>0.80) while roughly doubling precision to 0.80--0.86. Consequently, Rulemapping accuracies (0.90--0.98) consistently exceed the best prompt-only configurations (0.78--0.90) across all LLMs. 95\% bootstrap confidence intervals based on 10,000 resamples were consistent with the observed precision and recall differences, with the full intervals reported in the accompanying GitHub repository.

\begin{table}[htb]
  \caption{Benchmark against the lay-consensus reference of \datasetA~(868 rows) by prompting method. Bold indicates per-method column maxima. Lists Precision ($P$), Recall ($R$), Accuracy (Acc).}
  \label{tab:bench-zufall-lay}
  \centering
  \small
  \sisetup{
    table-format=1.2,
    table-number-alignment = center,
    detect-weight = true
  }
  \setlength{\tabcolsep}{2.6pt}
  \renewcommand{\arraystretch}{1.10}
  \begin{tabular*}{\columnwidth}{@{\extracolsep{\fill}}
    l
    S S S
    S S S
    S S S
    @{}}
    \toprule
      & \multicolumn{3}{c}{Rulemapping} & \multicolumn{3}{c}{Long-Context} & \multicolumn{3}{c}{Zero-Context} \\
      LLM
        & \multicolumn{1}{c}{$P$}
        & \multicolumn{1}{c}{$R$}
        & \multicolumn{1}{c}{Acc}
        & \multicolumn{1}{c}{$P$}
        & \multicolumn{1}{c}{$R$}
        & \multicolumn{1}{c}{Acc}
        & \multicolumn{1}{c}{$P$}
        & \multicolumn{1}{c}{$R$}
        & \multicolumn{1}{c}{Acc} \\
    \cmidrule(r){1-1}\cmidrule(lr){2-4}\cmidrule(lr){5-7}\cmidrule(lr){8-10}
      Mistral Large & \bfseries 0.86 & 0.89 & \bfseries 0.98 & \bfseries 0.49 & 0.69 & \bfseries 0.90 & 0.35 & 0.89 & 0.84 \\
      GPT-4o        & 0.83 & 0.84 & 0.97 & 0.34 & 0.99 & 0.82 & \bfseries 0.37 & 0.91 & \bfseries 0.85 \\
      GPT-o1        & 0.80 & 0.82 & 0.97 & 0.36 & \bfseries 1.00 & 0.84 & 0.34 & \bfseries 1.00 & 0.82 \\
      Deepseek V3   & 0.48 & \bfseries 0.91 & 0.90 & 0.37 & 0.93 & 0.85 & 0.28 & \bfseries 1.00 & 0.77 \\
      OSS 120B      & 0.75 & 0.50 & 0.94 & 0.37 & 0.94 & 0.85 & 0.34 & 0.95 & 0.82 \\
      GPT-5 mini     & 0.73 & 0.14 & 0.92 & 0.29 & \bfseries 1.00 & 0.78 & 0.33 & 0.97 & 0.81 \\
    \bottomrule
  \end{tabular*}
\end{table}

Comparing baselines reveals that added context is insufficient to correct over-flagging. Long-Context prompting fails to consistently improve precision over Zero-Context and occasionally degrades recall substantially (e.g., Mistral Large drops from $0.89$ to $0.69$), whereas Rulemapping's structural constraints reliably enforce selectivity.

Three model groupings emerge only under Rulemapping: balanced performers (Mistral Large, GPT-4o, GPT-o1), an over-flagger (DeepSeek V3), and conservative under-flaggers (GPT-5 mini, OSS 120B). In contrast, prompt-only baselines mask these intrinsic differences, clustering all models in a high-recall/low-precision regime. This divergence indicates that while solely using LLMs elicits generic over-flagging, symbolic scaffolding exposes the distinct calibration profile of each underlying model.

\subsection{Alternative Expert-Subset Reference}

The expert subset in Table~\ref{tab:bench-zufall-expert} reveals a shift in model dynamics, most notably reversing the Rulemapping LLM ranking with GPT-4o outperforming Mistral Large (0.62 vs. 0.53 precision). This inversion is driven by false positive control rather than sensitivity: both models achieved identical recall (catching 8 of 9 positives), but GPT-4o generated fewer false alarms (5 vs. 7). However, these results must be interpreted with caution: with only 9 positive cases in the subset, a single classification error alters recall by approximately 11\% and precision by roughly 4\%, introducing substantial volatility.

\begin{table}[htb]
  \caption{Benchmark against the Expert annotated sample of \datasetA~(102 rows) by prompting method. Bold indicates per-method column maxima. Lists Precision ($P$), Recall ($R$), Accuracy (Acc).}
  \label{tab:bench-zufall-expert}
  \centering
  \small
  \sisetup{
    table-format=1.2,
    table-number-alignment = center,
    detect-weight = true
  }
  \setlength{\tabcolsep}{2.6pt}
  \renewcommand{\arraystretch}{1.10}
  \begin{tabular*}{\columnwidth}{@{\extracolsep{\fill}}
    l
    S S S
    S S S
    S S S
    @{}}
    \toprule
      & \multicolumn{3}{c}{Rulemapping} & \multicolumn{3}{c}{Long-Context} & \multicolumn{3}{c}{Zero-Context} \\
      LLM
        & \multicolumn{1}{c}{$P$}
        & \multicolumn{1}{c}{$R$}
        & \multicolumn{1}{c}{Acc}
        & \multicolumn{1}{c}{$P$}
        & \multicolumn{1}{c}{$R$}
        & \multicolumn{1}{c}{Acc}
        & \multicolumn{1}{c}{$P$}
        & \multicolumn{1}{c}{$R$}
        & \multicolumn{1}{c}{Acc} \\
    \cmidrule(r){1-1}\cmidrule(lr){2-4}\cmidrule(lr){5-7}\cmidrule(lr){8-10}
        GPT 4o        & {\bfseries 0.62} & 0.89 & {\bfseries 0.94} & 0.29 & {\bfseries 1.00} & 0.78 & {\bfseries 0.36} & {\bfseries 1.00} & {\bfseries 0.84} \\
        GPT o1        & 0.58 & 0.78 & 0.93 & 0.27 & {\bfseries 1.00} & 0.76 & 0.26 & {\bfseries 1.00} & 0.75 \\
        Deepseek V3   & 0.43 & {\bfseries 1.00} & 0.88 & 0.27 & 0.89 & 0.77 & 0.23 & {\bfseries 1.00} & 0.70 \\
        Mistral Large & 0.53 & 0.89 & 0.92 & 0.29 & 0.44 & {\bfseries 0.85} & 0.29 & 0.89 & 0.79 \\
        OSS 120B       & 0.60 & 0.33 & 0.92 & {\bfseries 0.35} & {\bfseries 1.00} & 0.83 & 0.31 & {\bfseries 1.00} & 0.80 \\
        GPT-5 mini     & 0.00 & 0.00 & 0.90 & 0.24 & {\bfseries 1.00} & 0.74 & 0.29 & {\bfseries 1.00} & 0.79 \\
      \bottomrule
    \end{tabular*}
\end{table}

Under these stricter conditions, calibration profiles diverge. GPT-5 mini exhibits extreme conservatism, predicting the positive class only once across the entire subset ($TP=0$, $FP=1$). This represents a failure of calibration, as the model's confidence thresholds were rarely met under the strict symbolic scaffold. Conversely, the prompt-only baselines show no such adaptation; they remain trapped in a high-recall/low-precision regime (e.g., Long-Context GPT-4o generates 22 false positives compared to Rulemapping's 5), suggesting they rely on broad toxicity heuristics that fail to align with the expert's precise legal definitions.

\section{Discussion}
\label{sec:discussion}

Constraining LLMs within deterministic statutory scaffolds like Rulemapping improves precision at a controlled recall cost. While LLM-only baselines achieve recall approaching 1.0 by over-flagging, Rulemapping's top configurations maintain recall between 0.82--0.89 while increasing precision to 0.80--0.86, compared to 0.34--0.49 (see Table~\ref{tab:bench-zufall-lay}) for direct prompting.

LLM-only approaches over-flag because pretraining objectives diverge from legal tasks. General hate-speech corpora condition models to detect semantic toxicity, which correlates only loosely with criminal liability. For instance, in the DeTox dataset, general toxicity labels outnumber illegality labels by roughly 9:1 for the same posts~\cite[p.~149]{demusDeToxComprehensiveDataset2022}. This imbalance causes a \enquote{scope drift} where models conflate moral offensiveness with legal illegality. By enforcing a specific logical decomposition, Rulemapping corrects for this distributional bias, ensuring that inference remains grounded in the narrower statutory text rather than the broader, toxicity-weighted distributions of the model's training data. A striking example in our data involved Holocaust denial: most baselines flagged these posts---sometimes on mere mention---based on their general toxicity or implicit knowledge of \S~130(3) StGB (the specific ban on Holocaust denial), even though the task was explicitly restricted to \S~130(1). Rulemapping prevents such drift by enforcing decomposition at each statutory element, ensuring inferences remain within the defined scope. This confirms that neuro-symbolic designs effectively curb the tendency of LLMs to conflate general offensiveness with specific illegality.

The distinct error profiles suggest a pragmatic division of labor for automated illegality assessment at an industrial scale---a necessity for platforms and Article~21 Settlement Bodies under the Digital Services Act. Since LLM-only prompting catches virtually all potential violations but suffers from low precision (e.g., Zero-Context with a capable LLM to minimise token usage), it functions effectively as a high-volume screener. Rulemapping then serves as a precise second-stage analyser to confirm punishability and reduce false positives before expert review.

While our evaluation focused on \S~130(1) StGB, these findings transfer to the broader civil law tradition due to the universality of the \textit{Subsumtion} method (syllogistic application of rules, see also Section~\ref{sec:method}). The neuro-symbolic scaffold automates this logic by separating the normative premise (the applicable legal norm operationalised as statutory elements) from the factual subsumption (characterising facts against those elements). Rulemapping imposes determinism even on a high-ambiguity criminal statute like the incitement to hatred in \S~130 StGB; this suggests potential for even greater reliability for domains where decision-making is rule-bound rather than discretionary, such as mass administrative processes.

The neuro-symbolic architecture also reduces infrastructure requirements. It acts as a performance multiplier, enabling high-quality legal automation even on restricted on-premise hardware. Mistral Large performs best on the lay-consensus reference (Table~\ref{tab:bench-zufall-lay}), surpassing larger systems like GPT-4o and GPT-o1, even though the Rulemap for \S~130(1) StGB was built and tested originally with GPT-4o. Rulemapping consistently upgraded results for all tested open-source models, squeezing frontier-grade performance from models that fit within the constraints of on-premise application required for strict data locality and auditability.

Absolute performance varies by annotation standard (Lay-Consensus or Expert-Subset). A likely contributor to the worse overall performance on the expert subset in Table~\ref{tab:bench-zufall-expert} compared to Table~\ref{tab:bench-zufall-lay} is case-mix: the expert subset contains a higher effective share of edge cases because we removed all items where the lay annotators disagreed for the lay-consensus reference. GPT-4o overtakes Mistral Large here by reducing false positives from 7 to 5. However, given that the expert subset contains only nine positive instances, these rank shifts should be viewed as indicative rather than definitive; minor variations in model behavior disproportionately impact the metrics here.

Although the Rulemap and prompts are fixed, LLM choice matters: we observed balanced high performers, over-flagging models, and under-flagging models, indicating that the neuro-symbolic approach is not agnostic to model calibration. Deepseek V3 in the expert subset (Table~\ref{tab:bench-zufall-expert}) exemplifies this sensitivity. Despite operating within the same symbolic scaffold, it achieved perfect recall (1.0) on the expert set but incurred the highest false positive count (12) of any neuro-symbolic configuration. While the symbolic architecture constrains the search space, it cannot fully override a model's intrinsic tendency toward over-flagging or under-flagging; the final precision-recall balance remains heavily dependent on the underlying model's reasoning profile.

Standard instruction-following models benefit most from external scaffolding. General-purpose models like Mistral Large outperformed reasoning-specialised models like GPT-o1, suggesting a broader architectural insight. Prior work indicates that reasoning models often favor self-directed chains of thought and may degrade under tightly prescriptive instructions~\cite{wuComparativeStudyReasoning2024}. By contrast, standard models appear to benefit from Rulemapping, which offloads the reasoning structure to the expert tree and allows the model to focus on local semantic interpretation.

Finally, the method addresses the knowledge-acquisition bottleneck by placing the system's architectural quality directly in the hands of legal professionals. Because the system performs only as well as the Rulemap that has been built, the visual interface is not merely a convenience but a quality-assurance mechanism: it allows domain experts to test and refine decision logic without programmer intermediation. This direct expert control enables a calibrated Rulemap to serve a dual purpose: as a deployable tool for practitioners and as a structured baseline for LLM-as-a-judge settings, where the explicitly defined scaffold provides verifiable reference logic against which purely neural models can be measured.

\section{Limitations \& Legal Considerations}
\label{sec:limitations}

This study addresses German-language content and \S~130(1) StGB specifically. However, as detailed in Section~\ref{sec:discussion}, the proposed architecture targets the universal syllogistic structure (\textit{Subsumtion}) of civil law, suggesting high transferability to other rule-based domains. While methodologically robust, empirical performance in new domains will naturally depend on the target jurisdiction, the availability of high-quality legal context for each element, and the language capabilities of the chosen LLM. The primary reference labels derive from lay annotations; while this is standard for hate-speech benchmarks, it approximates common understanding rather than the definitive judicial doctrine found in court rulings.

Rulemapping's reliance on strict Boolean composition presents a structural trade-off. While this brittleness is a design feature for legal defensibility---ensuring no mandatory element is skipped---it renders the system sensitive to hard error propagation: a single false negative in a mandatory leaf node defeats the entire classification. Unlike neural systems that absorb partial mistakes via soft weighting, this deterministic scaffold offers no recovery from leaf-level failures.

The reliance on handcrafted scaffolds limits scalability. Because both the Rulemaps and the element-specific context materials must be manually curated by domain experts, expanding the system to broader statutory corpora requires significant upfront investment.

Finally, while the architecture increases transparency, it does not eliminate bias; it relocates it from inference to ex ante design choices---tree construction and context selection---making system quality dependent on the correctness and completeness of the expert-defined scaffold~\cite{prussAIJurisprudenceLarge2025}. In short, the system remains strictly bound by the competence of its designer.

The experiments used a proprietary commercial implementation; however, the logic-tree methodology is implementation-agnostic. Inference was performed using a certified GDPR-compliant cloud provider under a zero-retention policy and Data Processing Agreement. To facilitate reproducibility while preventing misuse, all classification outputs are released in the repository referenced in Section~\ref{sec:results}, providing post IDs and generated labels. Due to the sensitive nature of the dataset, which contains potential hate speech, the raw post texts cannot be publicly redistributed. Researchers must obtain raw data directly from the original dataset~\cite{zufallLegalApproachHate2022} authors.

%The author is an employee of Rulemapping Group, which is the company developing products with the Rulemapping method. This research was not part of the work there nor was it sponsored by the company.

\section{Conclusion}
\label{sec:conclusion}

The automation of legal reasoning confronts imperfect alternatives: symbolic systems offer transparency and precision but struggle with ambiguity, while neural systems handle natural language flexibly but lack verifiable determinism. Neither approach alone suffices. Our findings on hate speech classification under \S~130(1) StGB suggest that neuro-symbolic integration---specifically the Rulemapping architecture---addresses these limitations in the evaluated domain. By constraining neural components within a symbolic scaffold, the approach combines the linguistic flexibility of LLMs with the logical determinism of rule-based systems.

Our results indicate that architectural constraints significantly influence model performance. We found that, on the lay-consensus benchmark (Table~\ref{tab:bench-zufall-lay}), even mid-scale open-weight models (e.g., Mistral Large) outperformed frontier reasoning models (e.g., GPT-o1) when the former were constrained by a sound neuro-symbolic architecture. This does not displace the value of general scaling, but it shows that legal AI research can obtain performance gains by designing architectures that decompose tasks into auditable symbolic structure with narrowly scoped neural judgments. This is particularly relevant for public administration and legal departments where deployments require both strict adherence to statutory logic and on-premise operation for data locality, as the method enhances the utility of locally hostable models.

Consequently, transparency and precision reinforce each other. The logical decomposition enables both qualities through distinct mechanisms: the visual symbolic scaffolding ensures auditability, while the architectural constraints mitigate scope drift, thereby improving precision. This suggests that in statutory classification, transparency and performance need not be competing objectives; instead, they can align under a structure that reflects the domain's logical requirements.

Finally, this approach exemplifies how legal automation can preserve human expertise and control. Law remains a domain of human judgment and institutional authority, grounded in normative validity rather than statistical probability. Rulemapping reflects this by delegating narrowly scoped assessments to neural components while reserving overall reasoning authority for deterministic, expert-authored structures. The system augments human experts, enabling them to scale their own legal assessment capabilities while preserving their judgment authority.

\section{Future Work}
\label{sec:future-work}

Key research directions include domain validation, architectural refinement, and context automation. Validating the architecture on larger datasets across differing fields of law would assess generalisability beyond \S~130(1) StGB. Automating scaffold drafting---using LLMs to generate preliminary versions from human-authored exemplars for expert verification---could reduce the upfront investment required for new statutory domains while maintaining normative validity.

Three complementary strategies address the Boolean classification limitation identified in Section~\ref{sec:limitations}. First, expanding benchmarking to evaluate detailed reasoning explanations generated by LLMs, rather than relying solely on binary classifications, would enable the diagnosis of specific inferential failures driving incorrect results. Second, replacing purely Boolean leaf outputs with calibrated probabilistic scores offers a path to graded confidence. LLM-as-a-Judge approaches employing self-consistency and majority voting across multiple models could reflect uncertainty while maintaining compositional structure. Third, granularising neural leaves by incorporating computational argumentation and case-based reasoning would allow borderline cases to be represented with greater symbolic nuance, keeping the reasoning logic transparent and modifiable by domain experts.

Addressing manual context curation without reintroducing the opacity inherent in standard retrieval requires alternative architectures. GraphRAG~\cite{edgeLocalGlobalGraph2025} offers one promising direction: unlike unstructured vector search, it leverages explicit structural links between statutory elements and legal commentary to enable dynamic yet verifiable context curation. This maintains the expert's ability to inspect and verify the sources informing each neural judgment, ensuring that even dynamic context retrieval remains grounded in the transparent statutory scaffold described in Section~\ref{sec:neuro-symbolic}.

Beyond technical improvements, this structural decomposition offers a potential analytical lens for comparative law. By formalising the logical decomposition of statutes, Rulemapping may reveal how different jurisdictions operationalise concepts such as \enquote{incitement} or \enquote{disturbance of the public peace}, offering methodological tools for analysing the structural logic of law across legal systems.

The field is currently in an exploration phase regarding neuro-symbolic AI, and as Kautz~\cite{kautzThirdAISummer2022} outlines, there are many architectural frameworks to be explored, of which we have only touched upon one---one that we deem particularly well-suited for the demands of automated legal decision-making. While the most promising long-term direction is arguably a synthesis comparable to the cooperation between \enquote{System 1} (fast, intuitive pattern-matching) and \enquote{System 2} (slow, verifiable logic) described by Kahneman~\cite{kahnemanThinkingFastSlow2011}, distinct pathways exist to achieve it. Rulemapping operationalises this duality extrinsically rather than intrinsically: it effectively chains the two systems by subjecting neural \enquote{System 1} judgments to a symbolic \enquote{System 2} controller. Future work will determine whether this extrinsic orchestration offers the superior paradigm for law, or if genuine intrinsic reasoning capabilities can eventually be successfully embedded within LLMs (or their successors) themselves.

\bibliographystyle{ACM-Reference-Format} 
\bibliography{library_bibtex}

@article{anstotz130StGBVolksverhetzung2025,
  title = {\S{} 130 {{StGB}} - {{Volksverhetzung}}},
  shorttitle = {{{M\"uKo}}},
  author = {Anst{\"o}tz, Stephan},
  editor = {Erb, Volker and Sch{\"a}fer, J{\"u}rgen},
  year = 2025,
  journal = {M\"unchener Kommentar zum StGB},
  edition = {5},
  publisher = {C.H. Beck},
  address = {M\"unchen},
  urldate = {2025-10-08},
  isbn = {978-3-406-81310-8}
}

@article{ariaiNaturalLanguageProcessing2025,
  title = {Natural {{Language Processing}} for the {{Legal Domain}}: {{A Survey}} of {{Tasks}}, {{Datasets}}, {{Models}}, and {{Challenges}}},
  shorttitle = {Natural {{Language Processing}} for the {{Legal Domain}}},
  author = {Ariai, Farid and Mackenzie, Joel and Demartini, Gianluca},
  year = 2025,
  month = dec,
  journal = {ACM Comput. Surv.},
  volume = {58},
  number = {6},
  pages = {163:1--163:37},
  issn = {0360-0300},
  doi = {10.1145/3777009},
  urldate = {2026-01-13}
}

@article{bench-caponRulebasedRepresentationOpen1988,
  title = {Towards a Rule-Based Representation of Open Texture in Law},
  author = {{Bench-Capon}, Trevor and Sergot, M.},
  year = 1988,
  journal = {Computing Power and Legal Language},
  pages = {39--60}
}

@misc{bommaritoKL3MDataProject2025,
  type = {{{SSRN Scholarly Paper}}},
  title = {The {{KL3M Data Project}}: {{Copyright-Clean Training Resources}} for {{Large Language Models}}},
  shorttitle = {The {{KL3M Data Project}}},
  author = {Bommarito, Michael James and Bommarito, Jillian and Katz, Daniel Martin},
  year = 2025,
  month = apr,
  number = {5211933},
  eprint = {5211933},
  publisher = {Social Science Research Network},
  address = {Rochester, NY},
  doi = {10.2139/ssrn.5211933},
  urldate = {2025-11-14},
  archiveprefix = {Social Science Research Network},
  langid = {english}
}

@book{breidenbachRechtshandbuchLegalTech2021,
  title = {{Rechtshandbuch Legal Tech}},
  author = {Breidenbach, Stephan and Glatz, Florian and Braegelmann, Tom and Caba, Philipp and Dietzen, Alexandra},
  year = 2021,
  edition = {2},
  publisher = {C.H. Beck},
  address = {M\"unchen},
  urldate = {2025-10-29},
  isbn = {978-3-406-73830 2},
  langid = {german}
}

@article{conradsJuristischeProblemlosungMit2025,
  title = {Juristische {{Probleml\"osung}} Mit {{KI}} -- {{Leistung}} Und {{Grenzen}} Gro\ss er {{Sprachmodelle}}},
  author = {Conrads, Markus and Schweitzer, Sascha},
  year = 2025,
  journal = {Neue Juristische Wochenschrift},
  number = {40},
  pages = {2888--2891},
  issn = {0341-1915},
  url = {https://beck-online.beck.de/Bcid/Y-300-Z-NJW-B-2025-S-2888-N-1},
  urldate = {2025-10-06}
}

@article{dahlLargeLegalFictions2024,
  title = {Large {{Legal Fictions}}: {{Profiling Legal Hallucinations}} in {{Large Language Models}}},
  shorttitle = {Large {{Legal Fictions}}},
  author = {Dahl, Matthew and Magesh, Varun and Suzgun, Mirac and Ho, Daniel E},
  year = 2024,
  month = jan,
  journal = {Journal of Legal Analysis},
  volume = {16},
  number = {1},
  pages = {64--93},
  issn = {2161-7201},
  doi = {10.1093/jla/laae003},
  urldate = {2025-11-14}
}

@misc{deepseek-aiDeepSeekV3TechnicalReport2025,
  title = {{{DeepSeek-V3 Technical Report}}},
  year = 2025,
  month = feb,
  number = {arXiv:2412.19437},
  eprint = {2412.19437},
  primaryclass = {cs},
  publisher = {arXiv},
  doi = {10.48550/arXiv.2412.19437},
  urldate = {2025-10-28},
  archiveprefix = {arXiv},
  author = {DeepSeek-AI and others}
}

@inproceedings{demusDeToxComprehensiveDataset2022,
  title = {{{DeTox}}: {{A Comprehensive Dataset}} for {{German Offensive Language}} and {{Conversation Analysis}}},
  shorttitle = {{{DeTox}}},
  booktitle = {Proceedings of the {{Sixth Workshop}} on {{Online Abuse}} and {{Harms}} ({{WOAH}})},
  editor = {Narang, Kanika and Mostafazadeh Davani, Aida and Mathias, Lambert and Vidgen, Bertie and Talat, Zeerak},
  year = 2022,
  month = jul,
  pages = {143--153},
  publisher = {Association for Computational Linguistics},
  address = {Seattle, Washington (Hybrid)},
  doi = {10.18653/v1/2022.woah-1.14},
  urldate = {2025-07-02},
  author = {Demus, Christoph and others}
}

@article{djeffalCommentaryCommentariesWissenschaftsrat2014,
  title = {A {{Commentary}} on {{Commentaries}}: {{The Wissenschaftsrat}} on {{Legal Commentaries}} and {{Beyond}}},
  shorttitle = {A {{Commentary}} on {{Commentaries}}},
  author = {Djeffal, Christian},
  year = 2014,
  month = jun,
  journal = {Verfassungsblog},
  publisher = {Verfassungsblog},
  issn = {2366-7044},
  doi = {10.17176/20181005-163652-0},
  urldate = {2025-10-19},
  copyright = {CC BY-NC-ND 4.0},
  langid = {english}
}

@article{ederMakingSystemicRisk2024,
  title = {Making {{Systemic Risk Assessments Work}}: {{How}} the {{DSA Creates}} a {{Virtuous Loop}} to {{Address}} the {{Societal Harms}} of {{Content Moderation}}},
  shorttitle = {Making {{Systemic Risk Assessments Work}}},
  author = {Eder, Niklas},
  year = 2024,
  month = oct,
  journal = {German Law Journal},
  volume = {25},
  number = {7},
  pages = {1197--1218},
  issn = {2071-8322},
  doi = {10.1017/glj.2024.24},
  urldate = {2025-12-08},
  langid = {english}
}

@misc{edgeLocalGlobalGraph2025,
  title = {From {{Local}} to {{Global}}: {{A Graph RAG Approach}} to {{Query-Focused Summarization}}},
  shorttitle = {From {{Local}} to {{Global}}},
  year = 2025,
  month = feb,
  number = {arXiv:2404.16130},
  eprint = {2404.16130},
  primaryclass = {cs},
  publisher = {arXiv},
  doi = {10.48550/arXiv.2404.16130},
  urldate = {2026-01-22},
  archiveprefix = {arXiv},
  author = {Edge, Darren and others}
}

@misc{fanLEXamBenchmarkingLegal2025,
  title = {{{LEXam}}: {{Benchmarking Legal Reasoning}} on 340 {{Law Exams}}},
  shorttitle = {{{LEXam}}},
  year = 2025,
  month = may,
  number = {arXiv:2505.12864},
  eprint = {2505.12864},
  primaryclass = {cs},
  publisher = {arXiv},
  doi = {10.48550/arXiv.2505.12864},
  urldate = {2025-07-01},
  archiveprefix = {arXiv},
  author = {Fan, Yu and others}
}

@article{furstChallengesOpportunitiesVisual2025,
  title = {Challenges and {{Opportunities}} for {{Visual Analytics}} in {{Jurisprudence}}},
  author = {F{\"u}rst, Daniel and {El-Assady}, Mennatallah and Keim, Daniel A. and Fischer, Maximilian T.},
  year = 2025,
  month = nov,
  journal = {Artificial Intelligence and Law},
  issn = {1572-8382},
  doi = {10.1007/s10506-025-09494-2},
  urldate = {2025-12-09},
  langid = {english}
}

@article{guittonChallengeOpentextureLaw2025,
  title = {The Challenge of Open-Texture in Law},
  author = {Guitton, Clement and {Tam{\`o}-Larrieux}, Aurelia and Mayer, Simon and {van Dijck}, Gijs},
  year = 2025,
  month = jun,
  journal = {Artificial Intelligence and Law},
  volume = {33},
  number = {2},
  pages = {405--435},
  issn = {1572-8382},
  doi = {10.1007/s10506-024-09390-1},
  urldate = {2025-09-21},
  langid = {english}
}

@inproceedings{guoInvestigationLargeLanguage2023,
  title = {An Investigation of Large Language Models for Real-World Hate Speech Detection},
  booktitle = {2023 {{International Conference}} on {{Machine Learning}} and {{Applications}} ({{ICMLA}})},
  author = {Guo, Keyan and Hu, Alexander and Mu, Jaden and Shi, Ziheng and Zhao, Ziming and Vishwamitra, Nishant and Hu, Hongxin},
  year = 2023,
  pages = {1568--1573},
  publisher = {IEEE},
  url = {https://ieeexplore.ieee.org/abstract/document/10459901/},
  urldate = {2025-12-11}
}

@article{hamannTransparenzJustizStagnation2021,
  title = {{Transparenz der Justiz: Stagnation seit 50 Jahren}},
  shorttitle = {{Transparenz der Justiz}},
  author = {Hamann, Hanjo},
  year = 2021,
  month = jul,
  journal = {Legal Tribune Online},
  url = {https://www.lto.de/persistent/a_id/45370},
  urldate = {2025-12-11},
  langid = {ngerman}
}

@incollection{hartConceptLaw1961,
  title = {The {{Concept}} of {{Law}}},
  booktitle = {The {{Concept}} of {{Law}}},
  author = {Hart, H. L. A.},
  year = 1961,
  publisher = {Oxford University Press},
  doi = {10.2307/2217213},
  urldate = {2025-09-21},
  isbn = {978-0-19-181137-1},
  langid = {american}
}

@book{kahnemanThinkingFastSlow2011,
  title = {Thinking, {{Fast}} and {{Slow}}},
  author = {Kahneman, Daniel},
  year = 2011,
  month = oct,
  publisher = {{Farrar, Straus and Giroux}},
  googlebooks = {ZuKTvERuPG8C},
  isbn = {978-1-4299-6935-2},
  langid = {english}
}

@misc{kantEquitableAccessJustice2024,
  title = {Equitable {{Access}} to {{Justice}}: {{Logical LLMs Show Promise}}},
  shorttitle = {Equitable {{Access}} to {{Justice}}},
  author = {Kant, Manuj and Kant, Manav and Nabi, Marzieh and Carlson, Preston and Ma, Megan},
  year = 2024,
  month = oct,
  number = {arXiv:2410.09904},
  eprint = {2410.09904},
  primaryclass = {cs},
  publisher = {arXiv},
  doi = {10.48550/arXiv.2410.09904},
  urldate = {2025-09-15},
  archiveprefix = {arXiv}
}

@misc{kantRobustLegalReasoning2025,
  title = {Towards {{Robust Legal Reasoning}}: {{Harnessing Logical LLMs}} in {{Law}}},
  shorttitle = {Towards {{Robust Legal Reasoning}}},
  author = {Kant, Manuj and Nabi, Sareh and Kant, Manav and Scharrer, Roland and Ma, Megan and Nabi, Marzieh},
  year = 2025,
  month = feb,
  number = {arXiv:2502.17638},
  eprint = {2502.17638},
  primaryclass = {cs},
  publisher = {arXiv},
  doi = {10.48550/arXiv.2502.17638},
  urldate = {2025-08-12},
  archiveprefix = {arXiv}
}

@article{kautzThirdAISummer2022,
  title = {The {{Third AI Summer}}: {{AAAI Robert S}}. {{Engelmore Memorial Lecture}}},
  shorttitle = {The {{Third AI Summer}}},
  author = {Kautz, Henry},
  year = 2022,
  month = mar,
  journal = {AI Magazine},
  volume = {43},
  number = {1},
  pages = {105--125},
  issn = {2371-9621},
  doi = {10.1002/aaai.12036},
  urldate = {2025-09-15},
  copyright = {Copyright (c) 2022 AI Magazine},
  langid = {english}
}

@book{larenzMethodenlehreRechtswissenschaft1995,
  title = {Methodenlehre Der {{Rechtswissenschaft}}},
  author = {Larenz, Karl and Canaris, Claus-Wilhelm},
  year = 1995,
  series = {Springer-{{Lehrbuch}}},
  publisher = {Springer},
  address = {Berlin, Heidelberg},
  doi = {10.1007/978-3-662-08709-1},
  urldate = {2025-09-21},
  copyright = {http://www.springer.com/tdm},
  isbn = {978-3-540-59086-6 978-3-662-08709-1}
}

@inproceedings{ludwigConditioningLargeLanguage2025,
  title = {Conditioning {{Large Language Models}} on {{Legal Systems}}? {{Detecting Punishable Hate Speech}}},
  shorttitle = {Conditioning {{Large Language Models}} on {{Legal Systems}}?},
  booktitle = {Proceedings of the 21st {{Conference}} on {{Natural Language Processing}} ({{KONVENS}} 2025): {{Long}} and {{Short Papers}}},
  author = {Ludwig, Florian and Zesch, Torsten and Zufall, Frederike},
  editor = {Wartena, Christian and Heid, Ulrich},
  year = 2025,
  month = sep,
  pages = {154--167},
  publisher = {HsH Applied Academics},
  address = {Hannover, Germany},
  url = {https://aclanthology.org/2025.konvens-1.14/},
  urldate = {2026-01-13}
}

@article{mageshHallucinationFreeAssessingReliability2025,
  title = {Hallucination-{{Free}}? {{Assessing}} the {{Reliability}} of {{Leading AI Legal Research Tools}}},
  shorttitle = {Hallucination-{{Free}}?},
  author = {Magesh, Varun and Surani, Faiz and Dahl, Matthew and Suzgun, Mirac and Manning, Christopher D. and Ho, Daniel E.},
  year = 2025,
  journal = {Journal of Empirical Legal Studies},
  volume = {22},
  number = {2},
  pages = {216--242},
  issn = {1740-1461},
  doi = {10.1111/jels.12413},
  urldate = {2025-11-17},
  copyright = {\copyright{} 2025 The Author(s). Journal of Empirical Legal Studies published by Cornell Law School and Wiley Periodicals LLC.},
  langid = {english}
}

@article{mclachlanLawmapsEnablingLegal2022,
  title = {Lawmaps: Enabling Legal {{AI}} Development through Visualisation of the Implicit Structure of Legislation and Lawyerly Process},
  shorttitle = {Lawmaps},
  author = {McLachlan, Scott and Kyrimi, Evangelia and Dube, Kudakwashe and Fenton, Norman and Webley, Lisa C.},
  year = 2022,
  month = mar,
  journal = {Artificial Intelligence and Law},
  volume = {31},
  number = {1},
  pages = {169--194},
  issn = {1572-8382},
  doi = {10.1007/s10506-021-09298-0},
  urldate = {2025-11-25},
  langid = {english}
}

@article{mclachlanVisualisationLawLegal2021,
  title = {Visualisation of Law and Legal {{Process}}: {{An}} Opportunity Missed},
  shorttitle = {Visualisation of Law and Legal {{Process}}},
  author = {McLachlan, Scott and Webley, Lisa C},
  year = 2021,
  month = jul,
  journal = {Information Visualization},
  volume = {20},
  number = {2-3},
  pages = {192--204},
  issn = {1473-8716, 1473-8724},
  doi = {10.1177/14738716211012608},
  urldate = {2025-12-09},
  langid = {english}
}

@inproceedings{medvedevaLegalJudgmentPrediction2023,
  title = {Legal {{Judgment Prediction}}: {{If You Are Going}} to {{Do It}}, {{Do It Right}}},
  shorttitle = {Legal {{Judgment Prediction}}},
  booktitle = {Proceedings of the {{Natural Legal Language Processing Workshop}} 2023},
  author = {Medvedeva, Masha and Mcbride, Pauline},
  editor = {{Preo{\textcommabelow t}iuc-Pietro}, Daniel and Goanta, Catalina and Chalkidis, Ilias and Barrett, Leslie and Spanakis, Gerasimos and Aletras, Nikolaos},
  year = 2023,
  month = dec,
  pages = {73--84},
  publisher = {Association for Computational Linguistics},
  address = {Singapore},
  doi = {10.18653/v1/2023.nllp-1.9},
  urldate = {2025-11-14}
}

@inproceedings{mikolovDistributedRepresentationsWords2013,
  title = {Distributed {{Representations}} of {{Words}} and {{Phrases}} and Their {{Compositionality}}},
  booktitle = {Advances in {{Neural Information Processing Systems}}},
  author = {Mikolov, Tomas and Sutskever, Ilya and Chen, Kai and Corrado, Greg S and Dean, Jeff},
  year = 2013,
  volume = {26},
  publisher = {Curran Associates, Inc.},
  url = {https://papers.nips.cc/paper_files/paper/2013/hash/9aa42b31882ec039965f3c4923ce901b-Abstract.html},
  urldate = {2025-11-14}
}

@misc{mistralaiteamPixtralLargeMistral2024,
  title = {Pixtral {{Large}} and {{Mistral Large}} 2.1},
  author = {{Mistral AI Team}},
  year = 2024,
  month = nov,
  url = {https://mistral.ai/news/pixtral-large},
  urldate = {2025-12-17},
  langid = {english}
}

@article{mitschVolksverhetzungGegenDeutsche2011,
  title = {{Volksverhetzung gegen Deutsche}},
  author = {Mitsch, Wolfgang},
  year = 2011,
  journal = {Juristische Rundschau},
  volume = {2011},
  number = {9},
  pages = {380--382},
  doi = {10.1515/juru.2011.380},
  urldate = {2026-01-13},
  langid = {german}
}

@article{molesLogicProgrammingAssessment1991,
  title = {Logic Programming - {{An Assessment Of Its Potential For Artificial Intelligence Applications In Law}}},
  author = {Moles, Robert N},
  year = 1991,
  journal = {Journal of Law and Information Science},
  volume = {Vol 2 No. 2},
  pages = {137--164},
  url = {https://heinonline.org/HOL/P?h=hein.journals/jlinfos2&i=145},
  langid = {english}
}

@incollection{mumfordReasoningLegalCases2022,
  title = {Reasoning with {{Legal Cases}}: {{A Hybrid ADF-ML Approach}}},
  shorttitle = {Reasoning with {{Legal Cases}}},
  booktitle = {Legal {{Knowledge}} and {{Information Systems}}},
  author = {Mumford, Jack and Atkinson, Katie and {Bench-Capon}, Trevor},
  year = 2022,
  pages = {93--102},
  publisher = {IOS Press},
  doi = {10.3233/FAIA220452},
  urldate = {2025-10-15},
  langid = {english}
}

@inproceedings{nguyenLawBinaryTree2022,
  title = {Law to {{Binary Tree}} -- {{An Formal Interpretation}} of {{Legal Natural Language}}},
  booktitle = {Proceedings of the {{International Workshop}} on {{Methodologies}} for {{Translating Legal Norms}} into {{Formal Representations}}},
  author = {Nguyen, Ha-Thanh and Tran, Vu and Le, Ngoc-Cam and Le, Thi-Thuy and Nguyen, Quang-Huy and Nguyen, Le-Minh and Satoh, Ken},
  year = 2022,
  month = dec,
  eprint = {2212.08335},
  primaryclass = {cs},
  publisher = {arXiv},
  address = {Saarbr\"ucken},
  doi = {10.48550/arXiv.2212.08335},
  urldate = {2025-11-09},
  archiveprefix = {arXiv}
}

@misc{openaiGPT4oSystemCard2024,
  title = {{{GPT-4o System Card}}},
  year = 2024,
  month = oct,
  number = {arXiv:2410.21276},
  eprint = {2410.21276},
  primaryclass = {cs},
  publisher = {arXiv},
  doi = {10.48550/arXiv.2410.21276},
  urldate = {2025-10-28},
  archiveprefix = {arXiv},
  author = {OpenAI and others}
}

@misc{openaiGptoss120bGptoss20bModel2025,
  title = {Gpt-Oss-120b \& Gpt-Oss-20b {{Model Card}}},
  year = 2025,
  month = aug,
  number = {arXiv:2508.10925},
  eprint = {2508.10925},
  primaryclass = {cs},
  publisher = {arXiv},
  doi = {10.48550/arXiv.2508.10925},
  urldate = {2026-01-13},
  archiveprefix = {arXiv},
  author = {OpenAI and others}
}

@misc{openaiOpenAIO1System2024,
  title = {{{OpenAI}} O1 {{System Card}}},
  year = 2024,
  month = dec,
  number = {arXiv:2412.16720},
  eprint = {2412.16720},
  primaryclass = {cs},
  publisher = {arXiv},
  doi = {10.48550/arXiv.2412.16720},
  urldate = {2025-10-28},
  archiveprefix = {arXiv},
  author = {OpenAI and others}
}

@inproceedings{paniguttiRoleExplainableAI2023,
  title = {The Role of Explainable {{AI}} in the Context of the {{AI Act}}},
  booktitle = {Proceedings of the 2023 {{ACM Conference}} on {{Fairness}}, {{Accountability}}, and {{Transparency}}},
  year = 2023,
  month = jun,
  series = {{{FAccT}} '23},
  pages = {1139--1150},
  publisher = {Association for Computing Machinery},
  address = {New York, NY, USA},
  doi = {10.1145/3593013.3594069},
  urldate = {2025-11-14},
  isbn = {979-8-4007-0192-4},
  author = {Panigutti, Cecilia and others}
}

@article{prussAIJurisprudenceLarge2025,
  title = {Against {{AI Jurisprudence}}: {{Large Language Models}} and the {{False Promises}} of {{Empirical Judging}}},
  shorttitle = {Against {{AI Jurisprudence}}},
  author = {Pruss, Dasha and Allen, Jessie},
  year = 2025,
  month = oct,
  journal = {Proceedings of the AAAI/ACM Conference on AI, Ethics, and Society},
  volume = {8},
  number = {3},
  pages = {2055--2066},
  issn = {3065-8365},
  doi = {10.1609/aies.v8i3.36695},
  urldate = {2026-01-13},
  copyright = {Copyright (c) 2025 Association for the Advancement of Artificial Intelligence},
  langid = {english}
}

@article{reedAraucariaSoftwareArgument2004,
  title = {Araucaria: Software for Argument Analysis, Diagramming and Representation},
  shorttitle = {{{ARAUCARIA}}},
  author = {Reed, Chris and Rowe, Glenn},
  year = 2004,
  month = dec,
  journal = {International Journal on Artificial Intelligence Tools},
  volume = {13},
  number = {04},
  pages = {961--979},
  issn = {0218-2130, 1793-6349},
  doi = {10.1142/S0218213004001922},
  urldate = {2025-11-25},
  langid = {english}
}

@incollection{satohPROLEGPracticalLegal2023,
  title = {{{PROLEG}}: {{Practical Legal Reasoning System}}},
  booktitle = {Prolog: {{The Next}} 50 {{Years}}},
  author = {Satoh, Ken},
  year = 2023,
  series = {Lecture {{Notes}} in {{Computer Science}}},
  number = {13900},
  pages = {277--283},
  publisher = {Springer},
  doi = {10.1007/978-3-031-35254-6_23},
  isbn = {978-3-031-35253-9},
  langid = {english}
}

@article{sergotBritishNationalityAct1986,
  title = {The {{British Nationality Act}} as a Logic Program},
  author = {Sergot, M. J. and Sadri, F. and Kowalski, R. A. and Kriwaczek, F. and Hammond, P. and Cory, H. T.},
  year = 1986,
  month = may,
  journal = {Commun. ACM},
  volume = {29},
  number = {5},
  pages = {370--386},
  issn = {0001-0782},
  doi = {10.1145/5689.5920},
  urldate = {2025-10-31}
}

@article{shortliffeMycinKnowledgeBasedComputer1977,
  title = {Mycin: {{A Knowledge-Based Computer Program Applied}} to {{Infectious Diseases}}},
  shorttitle = {Mycin},
  author = {Shortliffe, Edward H.},
  year = 1977,
  month = oct,
  journal = {Proceedings of the Annual Symposium on Computer Application in Medical Care},
  pages = {66--69},
  issn = {0195-4210},
  url = {https://pmc.ncbi.nlm.nih.gov/articles/PMC2464549/},
  urldate = {2025-11-14}
}

@article{silverMasteringGameGo2016,
  title = {Mastering the Game of {{Go}} with Deep Neural Networks and Tree Search},
  year = 2016,
  month = jan,
  journal = {Nature},
  volume = {529},
  number = {7587},
  pages = {484--489},
  publisher = {Nature Publishing Group},
  issn = {1476-4687},
  doi = {10.1038/nature16961},
  urldate = {2025-11-09},
  copyright = {2016 Springer Nature Limited},
  langid = {english},
  author = {Silver, David and others}
}

@misc{singhOpenAIGPT5System2025,
  title = {{{OpenAI GPT-5 System Card}}},
  year = 2025,
  month = dec,
  number = {arXiv:2601.03267},
  eprint = {2601.03267},
  primaryclass = {cs},
  publisher = {arXiv},
  doi = {10.48550/arXiv.2601.03267},
  urldate = {2026-01-13},
  archiveprefix = {arXiv},
  author = {Singh, Aaditya and others}
}

@article{sternberg-lieben130StGBVolksverhetzung2025,
  title = {\S{} 130 {{StGB}} - {{Volksverhetzung}}},
  editor = {Eisele, J{\"o}rg and Bosch, Nikolaus and Kinzig, J{\"o}rg and Perron, Walter and Schittenhelm, Ulrike and Schuster, Frank and Steinberg, Georg and {Sternberg-Lieben}, Detlev and Wei{\ss}er, Bettina},
  year = 2025,
  journal = {T\"ubinger Kommentar Strafgesetzbuch},
  edition = {31},
  pages = {3555},
  publisher = {C.H. Beck},
  address = {M\"unchen},
  urldate = {2025-10-13},
  isbn = {978 3 406 80986 6},
  author = {Sternberg-Lieben, Detlev and others}
}

@article{streebTaskBasedVisualInteractive2022,
  title = {Task-{{Based Visual Interactive Modeling}}: {{Decision Trees}} and {{Rule-Based Classifiers}}},
  shorttitle = {Task-{{Based Visual Interactive Modeling}}},
  author = {Streeb, Dirk and Metz, Yannick and Schlegel, Udo and Schneider, Bruno and {El-Assady}, Mennatallah and Neth, Hansj{\"o}rg and Chen, Min and Keim, Daniel A.},
  year = 2022,
  month = sep,
  journal = {IEEE Transactions on Visualization and Computer Graphics},
  volume = {28},
  number = {9},
  pages = {3307--3323},
  issn = {1941-0506},
  doi = {10.1109/TVCG.2020.3045560},
  urldate = {2025-11-25}
}

@inproceedings{vaswaniAttentionAllYou2017,
  title = {Attention Is All You Need},
  booktitle = {Proceedings of the 31st {{International Conference}} on {{Neural Information Processing Systems}}},
  author = {Vaswani, Ashish and Shazeer, Noam and Parmar, Niki and Uszkoreit, Jakob and Jones, Llion and Gomez, Aidan N. and Kaiser, {\L}ukasz and Polosukhin, Illia},
  year = 2017,
  month = dec,
  series = {{{NIPS}}'17},
  pages = {6000--6010},
  publisher = {Curran Associates Inc.},
  address = {Red Hook, NY, USA},
  url = {https://dl.acm.org/doi/10.5555/3295222.3295349},
  urldate = {2026-01-13},
  isbn = {978-1-5108-6096-4}
}

@article{verheijDefLogLogicalInterpretation2003,
  title = {{{DefLog}}: On the {{Logical Interpretation}} of {{Prima Facie Justified Assumptions}}},
  shorttitle = {{{DefLog}}},
  author = {Verheij, Bart},
  year = 2003,
  month = jun,
  journal = {Journal of Logic and Computation},
  volume = {13},
  number = {3},
  pages = {319--346},
  issn = {1465-363X},
  doi = {10.1093/logcom/13.3.319},
  urldate = {2025-11-25}
}

@inproceedings{weiChainofthoughtPromptingElicits2022,
  title = {Chain-of-Thought Prompting Elicits Reasoning in Large Language Models},
  booktitle = {Proceedings of the 36th {{International Conference}} on {{Neural Information Processing Systems}}},
  author = {Wei, Jason and Wang, Xuezhi and Schuurmans, Dale and Bosma, Maarten and Ichter, Brian and Xia, Fei and Chi, Ed H. and Le, Quoc V. and Zhou, Denny},
  year = 2022,
  month = nov,
  series = {{{NIPS}} '22},
  pages = {24824--24837},
  publisher = {Curran Associates Inc.},
  address = {Red Hook, NY, USA},
  url = {https://proceedings.neurips.cc/paper_files/paper/2022/file/9d5609613524ecf4f15af0f7b31abca4-Paper-Conference.pdf},
  urldate = {2025-10-26},
  isbn = {978-1-7138-7108-8}
}

@misc{wuComparativeStudyReasoning2024,
  title = {A {{Comparative Study}} on {{Reasoning Patterns}} of {{OpenAI}}'s O1 {{Model}}},
  year = 2024,
  month = oct,
  number = {arXiv:2410.13639},
  eprint = {2410.13639},
  primaryclass = {cs},
  publisher = {arXiv},
  doi = {10.48550/arXiv.2410.13639},
  urldate = {2025-10-27},
  archiveprefix = {arXiv},
  author = {Wu, Siwei and others}
}

@inproceedings{zhangSyLeRFrameworkExplicit2025,
  title = {{{SyLeR}}: {{A Framework}} for {{Explicit Syllogistic Legal Reasoning}} in {{Large Language Models}}},
  shorttitle = {{{SyLeR}}},
  booktitle = {Proceedings of the 34th {{ACM International Conference}} on {{Information}} and {{Knowledge Management}}},
  author = {Zhang, Kepu and Yu, Weijie and Sun, Zhongxiang and Xu, Jun},
  year = 2025,
  month = nov,
  series = {{{CIKM}} '25},
  pages = {4117--4127},
  publisher = {Association for Computing Machinery},
  address = {New York, NY, USA},
  doi = {10.1145/3746252.3761120},
  urldate = {2025-11-25},
  isbn = {979-8-4007-2040-6}
}

@inproceedings{zufallLegalApproachHate2022,
  title = {A {{Legal Approach}} to {{Hate Speech}} -- {{Operationalizing}} the {{EU}}'s {{Legal Framework}} against the {{Expression}} of {{Hatred}} as an {{NLP Task}}},
  booktitle = {Proceedings of the {{Natural Legal Language Processing Workshop}} 2022},
  author = {Zufall, Frederike and Hamacher, Marius and Kloppenborg, Katharina and Zesch, Torsten},
  year = 2022,
  pages = {53--64},
  publisher = {Association for Computational Linguistics},
  address = {Abu Dhabi, United Arab Emirates (Hybrid)},
  doi = {10.18653/v1/2022.nllp-1.5},
  urldate = {2025-07-01},
  langid = {english}
}

\end{document}